\documentclass[preprint,aps,prd,nofootinbib]{revtex4}
\usepackage{amsmath}
\usepackage{latexsym}
\usepackage[pdfauthor={Diego N. Pelliccia},pdftitle=
{Electogenesis via primordial black holes},pdfsubject=
{photon mass}, pdfkeywords={primordial black holes, electric asymmetry, photon mass}]{hyperref}
\newcommand{\be}{\begin{equation}}
\newcommand{\ee}{\end{equation}}
\newcommand{\ba}{\begin{eqnarray}}
\newcommand{\ea}{\end{eqnarray}}
\newcommand{\bc}{\begin{center}}
\newcommand{\ec}{\end{center}}
\newcommand{\bi}{\bibitem}
\def\cstok#1{\leavevmode\thinspace\hbox{\vrule\vtop{\vbox{\hrule\kern1pt
\hbox{\vphantom{\tt/}\thinspace{\tt#1}\thinspace}}
\kern1pt\hrule}\vrule}\thinspace}
\begin{document}
\title{Electrogenesis via primordial black holes}
\author
{Diego N. Pelliccia$^{1,2}$
\thanks{Electronic address: diego.pelliccia@fe.infn.it}
}
\affiliation{ ${ }^{1}$Istituto Nazionale di
Fisica Nucleare, Sezione di Ferrara,\\
Via Saragat 1, 44100 Ferrara, Italy\\
${ }^{2}$Universit\`a di Ferrara,
Dipartimento di Fisica,\\
Via Saragat 1, 44100 Ferrara, Italy\\
}

\vspace{2cm}
\begin{abstract}

It was recently shown that the vanishing of the Coulomb field related to the electric charge of a black hole by virtue of a tiny but non-vanishing  photon mass can produce an electric asymmetry in the universe. This process can have place both at a cosmological early stage, through primordial black holes, and in contemporary universe, by means of supermassive black holes at galactic centers. Possible consequences of the latter case about the role of electrogenesis in the generation of magnetic fields coherent on galactic scale have been investigated. In this paper we will discuss the electrogenesis mediated by primordial black holes and analyze different realizations of the mechanism in this case.

\end{abstract}

\maketitle

\section{Introduction}

It is evident that the universe is asymmetric with respect to the baryonic charge. Constraints from Big Bang Nucleosynthesis (BBN) also leave open the possibility of an asymmetry in the leptonic charge. 
The baryon and lepton asymmetry are related with non--conservation of the baryonic and leptonic number, which are the corresponding charges of $U(1)$ global symmetries. In the standard model of particle physics these symmetries cannot be broken in the pertubative sector, by virtue of the couplings in the theory. In the non--perturbative sector, instead, there are topological non---trivial configurations of the non--Abelian gauge fields that lead to a violation of these numbers. The magnitudes of these effects are strongly suppressed and cannot generate the observed baryonic asymmetry $\beta_{B}= (N_{B}-N_{\overline{B}})/N_{\gamma}\approx 3 \times 10^{-10}$. 

In some models of baryogenesis primordial black holes (PBHs) are considered \cite{PBHB}. This opportunity relies on considering that the baryonic charge is related to a global symmetry in the Lagrangian of particle physics models. If the $B$ symmetry was an exact local symmetry, that is if the baryonic number was the charge of an unbroken $U(1)$ gauge theory, it would be mediated by massless vector bosons as in the standard electromagnetism. In this case it would be associated to a long range force.
Being a global symmetry, it could be mediated by a massive vector boson and associated with a finite range force. Constraints on the coupling of such a force were presented in the literature (also for an hypotethic leptonic photon). From tests of the validity of the equivalence principle one can get restrictive upper bounds on these coupling constants: $\alpha_{B}<10^{-44}$ and $\alpha_{L}<10^{-48}$ for the baryonic and leptonic interactions respectively \cite{LeeOkun}. 

It is well known that classical black holes (BHs) are characterized by only three quantities: their mass, their angular momentum (if rotating) and their electric charge (if charged). They cannot be endowed with any other field, as is stated by the "no hairs theorems" \cite{HE}. These properties of classical BHs make possible the baryonic asymmetry: when the baryonic charge carried by the particles which crossed the horizon disappears, after the complete evaporation of the BH, no remnant baryonic field is present to balance the asymmetry which originated outside the horizon at the crossing. The baryonic charge which entered the BH horizon is lost (in discussing the baryogenesis we will consider only the case of complete evaporation of electrically neutral Schwarzschild PBHs and consequent absence of any PBH remnant due to uknown quantum gravity effects).

The same is not possible if the considered charge is the electric charge. Being connected to an exact $U(1)$ gauge symmetry, every electric charge which crosses the horizon leaves its fingerprint in the Coulomb field of the BH. As the BH possesses an electric hair, no electric asymmetry due to the BH complete evaporation can be generated, as long as electric charge is supposed to be conserved, the latter condition being required by the gauge symmetry and demanded by experimental evidences.

Nevertheless in the following we will present some models of electrogenesis by means of PBHs. Our considerations are based on the relaxation of the requirement that the $U(1)$ gauge symmetry associated with electromagnetism must be unbroken. If this symmetry is not an exact local one, then the previous comments are no more valid and the generation of a cosmological electric asymmetry is possible. We will show two possibilities: the case of a photon with a constant mass, that is a global $U(1)$ symmetry for the electromagnetism, and the case of an early stage of the universe when the local symmetry can be spontaneously broken and later restored.

\section{Baryogenesis via Primordial black holes}

There are many different models for baryogenesis in the literature (for a review see e.g. \cite{DolgovMag}). Most of them employ some realization of the  three conditions proposed by Sakharov in 1967 \cite{Sakha}, which allow field theory to produce a baryonic asymmetry: 1) baryonic charge non--conservation, 2)  breaking of C and CP discrete symmetries and 3) deviation from thermal equilibrium.  Nevertheless some models do not require all the three conditions to be fullfilled. There are mechanisms which make use of PBHs and the properties related to the presence of their horizon.

The formation of a BH is a classical gravitational process, the occurrence of a BH is independent on the initial value of the mass which collapses in the gravitational potential and enters the horizon. Therefore it is assumed that this description works in the same way for supermassive BHs at the center of galaxies as well as for PBHs. About twenty confirmed candidates for astrophysical BHs are known, with mass in the range of stellar masses, 5-20 $M_{\odot}$, and three dozens of candidates for supermassive BHs, with mass in the range $10^{6}-10^{9.5} M_{\odot}$ \cite{Bilic}. There are several indirect methods for their detection \cite{Muller} and their existence is widely accepted among the astrophysical community.

The BH evaporation by the Hawking radiation \cite{Hawking} opens the chance for generating a cosmological baryonic asymmetry by virtue of $B$ non--conserving processes. The final fate of the BH is still not clear. The unknown role of quantum gravity effects when the BH mass equals the Planck mass does not allow one to predict if the BH leaves a Planck mass remnant behind it or if it completely disappears. In the latter case the information loss problem is still a matter of debate. After Bekenstein in 1972 showed the connection between the entropy and the BH area  \cite{Bekenstein}, it was  realized by Hawking in 1975 \cite{Hawking2} that the baryonic matter which crossed a BH horizon could then be emitted by quantum effects near the horizon in the form of thermal radiation indepent on the initial conditions and symmetric with respect to baryonic charge. In this way a BH formed by baryonic matter could disappear through its complete evaporation, leaving behind it only a charge symmetric radiation. This process might lead to a baryonic asymmetry in the universe, once charge symmetry breaking is also considered.

The baryonic charge conservation would forbid a baryon asymmetry in the case of non--interacting evaporated particles, due to the absence of a charge asymmetry in thermal equilibrium \cite{Touissant}. Particle interactions, on the other hand, allow the BH to evaporate asymmetrically even if the baryonic charge is conserved \cite{DolgovMag},\cite{DolgovHid} . 

This mechanism of baryogenesis concerns only the BHs whose mass is less than $M_{PBH}\leq M_{cr}=5 \times 10^{14} g$. BHs with greater values for their mass would have not completely evaporated yet, as will be shown later by the relation between the BH mass and its life-time. The origin of the BHs with a mass $M_{PBH}\leq 10^{-15}\,M_{\odot}$ cannot be the gravitational collapse of a star: their formation is supposed to occur at an early stage of the universe history, that is why they are called primordial black holes.         

There are strong constraints to the models of baryogenesis by means of PBH evaporation, because the energy release during the evaporation may affect the firm predictions on BBN, may distort the spectrum of Cosmic Microvawe Background Radiation (CMBR) or may affect the large scale structure formation (the constraints wil be shown in next sections). Therefore the initial density of the PBHs, which together with the PBH mass is a parameter of the models, cannot be too high, otherwise it would spoil the great agreement of particle physics predictions with the  cosmological data (see e.g. \cite{Muk}). These bounds render the models uneffective for the generation of the observed bayonic asymmetry.

Nevertheless, although PBHs cannot account for the value of $\beta_{B}$, neither can be detected by present gravitational antennas, they are not excluded by any theoretical principle. In some models of PBH remnants of the order of one Planck mass, in fact, the possible role of such PBHs as dark matter candidates is considered \cite{DolgovNN}. 

Baryogenesis in the presence of PBHs by means of charge asymmetric decays of heavy particles was considered in the literature. $B$ non-conserving processes and the departure from the equilibrium created by the BH, which acts as a source, grant the fulfilling of the three Sakharov's conditions. In different models of baryogenesis by means of heavy particle decays, the temperature may vary from the Planck mass $m_{Pl}=1.22 \times 10^{19}$ GeV to $O$(MeV), depending on the model. In the following we will adopt the system of units $c=h=1, G=1/m^{2}_{Pl}$.

\section{Photon mass and electrogenesis}

In contrast to the case of baryonic or leptonic charge, the electric charge of the universe is usually assumed to vanish exactly. There are stringent limits on the value of the charge-per-baryon ratio $\Delta$ at different universe's ages from observations: from anisotropies of cosmic rays and isotropy of the CMBR one has $|\Delta|<10^{-29}e$ at present time \cite{Delta1} and at last scattering $(z \approx 1089)$ \cite{Delta2} respectively, $e$ being the electric charge of the electron; from BBN constraints $|\Delta| \leq 10^{-32}e$ at $z \approx 4\times 10^{8}$ \cite{Delta3}.

Nevertheless several proposals for electrogenesis are in the literature, any of them consisting in a realization of a departure from the standard Maxwell electrodynamics. Electric current non--conservation and massive photons are the main ingredient to modify Maxwell equations. Although at the classical level the gauge invariance in electrodynamics implies the massless nature of photons, at the quantum level perturbatively renormalizable formulations of electrodynamics are possible, with the inclusion of massive photons. The limit of vanishing photon mass takes the modified theory smoothly over to massless QED \cite{Stu}. 

Theoretical studies of breaking of the electric current conservation have a long history \cite{Voloshin}, \cite{Voloshin2}. In different models of photon mass generation the authors considered temporarily breaking of the gauge symmetry in grand unification theories \cite{LangPi}, in spontaneous breaking with extra scalar fields \cite{DolgovSilk} or in higher dimensional theories \cite{Delta1},\cite{Nambu}; non exactness of the symmetry in theories with variable speed of light \cite{Landau}, in theories admitting variations of electromagnetic coupling constant $\alpha$ \cite{Shaw}, in brane world models \cite{Brane} and models with electron-positron oscillations \cite{Moha}; the topology of the non--perturbative sector of the electromagnetism with generalizations of the pioneer suggestion made by Schwinger in two dimensions \cite{Schwinger}, including the introduction of new fields \cite{Jackiw}; breaking of Lorentz invariance from string theories \cite{Bmota} or by the presence of Chern--Simons terms in the Lagrangian \cite{Carroll}; the role of gamma matrices and spinorial geometry in electrodynamics\cite{Giampi}. 

A non-null photon mass would lead to the observation of the wavelength dependence of the speed of light in free space\cite{Gammarev1}, the modifications of Coulomb's law \cite{Williams} and Ampere's law \cite{Chernikov}, the existence of longitudinal electromagnetic waves \cite{Bass}, the additional Yukawa potential of magnetic dipole fields \cite{BarrowFisc}, and the Faraday rotation of light from galaxies, with changes in polarization induced by galactic magnetic fields \cite{Carroll}. 

As well as investigations for massive photons spanned different arguments in physics (see e.g. \cite{Hollweg}), several related experiments have been performed \cite{Exp}. 
The experimental limits on photon mass are very stringent. They are based on observations of magnetic fields of celestial bodies or galactic magnetic fields (for a review see \cite{Gammarev}) and on laboratory experiments conducted with a rotating torsion balance. A robust upper bound is obtained from the measurement of the Jupiter magnetic field by the Pioneer-10 mission \cite{Davis}, giving 
\be
\label{m1}
m_{\gamma} < 6 \cdot 10^{-16} \textrm{eV}\qquad \textrm{or}\qquad \lambda_{\gamma}> 3 \cdot 10^{10} \quad\textrm{cm}\, , 
\ee
where $\lambda_{\gamma} \equiv 1/m_{\gamma}$ is the Compton wavelength of the photon.
In ref. \cite{Luo} a rotating torsion balance is used to detect the product of the photon mass squared and the ambient magnetic cosmic vector potential. Assuming the Coma galactic cluster values \cite{Asseo} for the magnetic fields and the distance ($\approx 0.2 \mu G$ and $\approx 4 \times 10^{22} m$ respectively), the experimental results give the following bound on the photon mass \cite{Luo}: 
\be
\label{m2}
m_{\gamma} < 0.7 \cdot 10^{-18} \textrm{eV}\qquad \textrm{or}\qquad \lambda_{\gamma}> 2 \cdot 10^{13} \quad\textrm{cm}\, . 
\ee
This bound is affected by statistical errors and it should be at least to orders of magnitude higher \cite{GNluo}, so although the tecnique is promising, limits on the photon mass coming from large scale magnetic fields are still more firm ones.
A much stronger bound can be obtained from the existence of magnetic fields coherent on galactic scale of few kpc \cite{Chibisov}:
\be
\label{m3}
m_{\gamma} < 10^{-27} \textrm{eV}\qquad \textrm{or}\qquad \lambda_{\gamma}>  10^{22} \quad\textrm{cm}\,.
\ee

If the photon acquires its mass through a spontaneous breaking of the $U(1)$ gauge symmetry, bound \eqref{m3} may be invalid \cite{Adelberger}, while bound \eqref{m1} still remains. 

The Gauss theorem implies that a closed universe must have zero net electric charge: if the charge would be non-zero, the imposition of the theorem  would lead to an inconsistency \cite{KimLee}. An open universe, on the other hand, may possess a non--zero electric charge, given by a homogeneous electric charge density $\sigma$.
From the Gauss law in the case of standard electrodynamics one has $div E=\sigma$, where E is the electric field. Solutions of this equations rise with distance, as can be easily seen. In the Friedmann--Robertson-Walker cosmology, where the Gauss law has the same form, this behavior destroys the homogeneity and the isotropy of the universe. Moreover the serious problem of infinitely rising energy density of the electric field is raised by these solutions with $E\sim x$. A small photon mass could remove this inconvenience. 

The cosmology of a charged universe is discussed in several papers \cite{DolgovMag},\cite{Delta1},\cite{Nambu},\cite{chargedcosmo},\cite{Siegel}, both with massless and massive photons. In this work we will show a possible model for the generation of the electric charge of the universe. In a previous paper \cite{DolgovMe} we presented a mechanism of electrogenesis which made use of the dynamical properties of the Coulomb field near BHs in the case of massive photons. We treated both the case of constant photon mass and the case of non-minimal coupling of the electromagnetic vector potential $A^{\mu}$ to the scalar curvature $R$, given by the term $\mathcal{L}_{int}= \xi R A^{\mu}A_{\mu}$ in the Lagrangian. Focusing on the possible role of this process in the generation of the galactic magnetic fields, we stressed the second case in the presence of supermassive BHs in the center of galaxies, surrounded by an electron--proton plasma. The larger mobility of the protons with respect to that of the electrons charges positively the BH. After a rapid vanishing of the Coulomb field of the BH, the uncompensated electric charge of electrons are repelled each from the other and create electric currents. 

This mechanism of electrogenesis works also for isolated PBHs, on which we focus in this paper.
To illustrate the mechanism of electric charge non-conservation, let us remind the following model \cite{Dolgovmodel}. 

Indipendently on its origin, if the photon mass is non zero, then the Maxwell equations of standard electrodynamics are modified in the Proca equations:
\be
\label{Pro}
\nabla_{\mu}F^{\mu\nu}+m^{2}_{\gamma}\,A^{\nu}=4\pi\,J^{\nu}  \,,
\ee
where $F^{\mu\nu}$ is the electromagnetic field $F^{\mu\nu} \equiv \nabla^{\mu}A^{\nu}-\nabla^{\nu}A^{\mu}$ and $J^{\nu}$ is the source term.

In flat space--time the Coulomb potential of standard electrodynamics is modified in the Yukawa one in the case of massive photons: 
\be
\label{Yukawa}
A_{t}=\frac{Q}{r} \quad \rightarrow \quad \frac{Q\,e^{-m_{\gamma} r}}{r} \,.
\ee
In the curved space--time outside a Schwarzschild BH, instead, the massive nature of the photon leads the Coulomb potential to be identically zero: 
\be
A_{t}=\frac{Q}{r} \quad \rightarrow A_{t}=0 \,.
\ee

This behavior can be seen by solving the Proca equation \eqref{Pro} in the Schwarzschild metric.
Consider a spherically symmetric uniformly charged shell with radius $R_{c}$, concentric with a spherically symmetric matter shell, whose gravitational radius is $R_{g}$. The charged shell is assumed to be outside the matter shell and having negligible mass.

Outside the matter shell the Proca equation for the Coulomb potential takes the form
\be
\label{Au}
\frac{1}{r^{2}}\left( r^{2} A^{\prime}_{t}\right)^{\prime} - \frac{m^{2}_{\gamma}\,r}{r-R_{g}}\,A_{t}=0 \,,
\ee
where prime means derivative with respect to the spatial coordinate.

Introducing the new function $q$ as an effective charge, $q\equiv r\,A_{t}$, and the new variables $y \equiv 2\, m_{\gamma}\,(r-R_{g})$ and $\mu \equiv m_{\gamma}\,R_{g}$, the Proca equation \eqref{Au} becomes:
\be
\label{Proca}
\frac{d^{2}q}{dy^{2}}-\left( \frac{1}{4} + \frac{\mu}{2\, y}\right)\,q=0 \, .
\ee  

The solution of this equation, which has the form of the Whittaker differential equation, can be expressed in terms of confluent hypergeometric functions \cite{Hyper}:
\be
\label{whit}
q(y)=C\,y\,e^{-y/2}\,\Phi(1+\mu/2,2,y) + B\,y\,e^{-y/2}\,\Psi(1+\mu/2,2,y) \,.
\ee
The coefficients $C$ and $B$ are constants and are determined by imposing appropriate boundary and junction conditions in order to ensure finiteness of the energy, namely the condition of vanishing of the field at infinity and of absence of singularity on the horizon $y=0$.

When the radius of the charged shell tends to the gravitational radius of the matter shell, that is $R_{c} \rightarrow R_{g}$, one obtains 
\be
\label{limit}
q \sim q_{0}\, \frac{R_{c}-R_{g}}{R_{c}}\,\Gamma(1+\mu/2) \,,
\ee
where $\Gamma$ is the usual gamma-function. 

From this behavior one can see that in the case of massive photons the Coulomb field vanishes as the charged shell approaches the Schwarzschild radius. The electric field disappears completely when the charged shell crosses the horizon, leading to an effective electric charge non-conservation, despite the formal current conservation. The vanishing of the electric field outside a Schwarzschild BH was demonstrated for a pointilike charge particle approaching the BH in the papers \cite{Dolgovmodel},\cite{LL}.

Contrary to the flat space-time situation (see e.g. \cite{Ticciati}), the limit $m_{\gamma} \rightarrow 0$ to the standard electrodynamics description can show a finite change of potential discontinous at the point $m_{\gamma}=0$ in the presence of a BH  \cite{Dolgovmodel}.

In wavefunction formalism, considering trasmission and reflection coefficients for the monopole component of the gauge field, it can be showed that the characteristic time of disappearance $\tau_{C}$ of the Coulomb field is inversely proportional to the photon mass \cite{Pawl}: 
\be
\label{tauC}
\tau_{C} \sim \frac{1}{m_{\gamma}}\,.
\ee

The disappearance of the electric hair of a BH and its evaporation lead to consider models in which electrogenesis occurs via PBHs.

\section{Primordial black holes}

As the characteristic of rotation of PBHs will not affect our considerations, although realistic BHs are supposed to rotate, we will ignore this further spreading in the parameter space.
 
Regardless of its origin, every classical BH is subjected to the quantum mechanical process of the Hawking evaporation, for which the following basic properties hold.

A BH with mass $M_{BH}$ radiates from its gravitational radius $R_{g}=2\,M_{BH}\,/m^{2}_{Pl}$ as a black body with temperature
\be
\label{BHT}
T_{BH}=\frac{m^{2}_{Pl}}{8\,\pi\,M_{BH}}\,,
\ee
that is the smaller the BH mass $M_{BH}$ is, the larger is the BH temperature $T_{BH}$.

The energy flux radiated by a body with temperature \eqref{BHT} is
\be
\label{BHen}
\dot{M}_{BH}=-\frac{\pi^{2}}{30}\,g_{*}\,T_{BH}^{4}\,\frac{4\,\pi\cdot4\,M_{BH}^{2}}{m_{Pl}^{4}} \approx -4\times 10^{-5}\,g_{*}\,\frac{m_{Pl}^{4}}{M_{BH}^{2}}\,,
\ee
where $g_{*}$ is the effective number of light particle species with $m<T_{BH}$.

After the particles are produced they propagate in the gravitational field of the BH and can even cross back the horizon. Their interaction depends on their mass and spin.

Integrating eq. \eqref{BHen} one obtains
\be
\label{BHCM}
M_{BH}(t)=M_{i}\,\left[1-(t-t_{i})/\tau_{BH}\right]^{1/3}\,,
\ee
where $M_{i}$ is the BH mass at initial time $t_{i}$ and
\be
\label{BHLT}
\tau_{BH}=\frac{1.33\,\times\,10^{5}}{g_{*}\,m_{Pl}}\,\left(\frac{M_{i}}{m_{Pl}}\right)^{3}\,
\ee
is the life-time of the BH with initial mass $M_{i}$.
The dependence of $g_{*}$ on $T_{BH}$ is neglected in \eqref{BHCM}, the obtained value is in good agreement with an order-of-magnitude estimate.

The parameters concerning the PBHs are still unknown, although more than thirty years have passed since the first proposals of PBHs for baryogenesis. Their initial mass $M_{i}$ and their initial number density, that is the fraction of the energy density they represent, are constrained from theoretical considerations and from indirect observational bounds. Whereas the first parameter dictates the life-time of the PBH and also its initial temperature, the other one determines whether PBH distribution dominates at some  universe's stage over radiation or matter distribution. A complete treatment of all possibilities of PBH dominance can be found in the literature  \cite{Kolb}. PBHs are interesting also because they can be used to constraints extensions of the standard model in particle physics \cite{Kh1,Kh2,Kh4,Kh15} and may give rise to supermassive BHs at galactic centers (with or without clustering) \cite{Kh12,Kh14}.

PBHs formed before an inflationary stage would be diluted away from the rapid expansion of the universe. The same phase of inflation has been considered in various models of generation of PBHs \cite{Bousso0}. The formation during inflation must take care of different inflationary models. In the standard scenario with one scalar field, whose dynamics describe an evolving cosmological constant, the probability of a universe with PBH formation is strongly suppressed with respect to the one of an inflating universe without PBH pair production. More precisely, the formation of PBHs with initial mass significantly larger than the Plank mass is very unlikely \cite{Bousso}. 

Moreover, as the magnitude of CMBR anisotropy is $\approx 10^{-5}$, the scalar field fluctuations should have a spectral shape such that it has an amplitude $\approx 10^{-5}$ on large scales, sharply increases by a factor $10^{4}$ on the mass scale of the PBHs and decreases again on smaller scales at the time of horizon crossing. This feature of the spectrum of fluctuations is difficult to obtain with a single scalar field. 
Alternatively, models with multiple scalar fields driven inflation or with the scalar potential with a plateau in the range mass of interest may lead to the PBH formation \cite{Bousso2}.

Their formation could also be caused by phase transitions \cite{Bousso3,Kh9}, bubble collisions \cite{Bousso4} and by the decay of cosmic loops \cite{Bousso5}. 

In the case of phase transitions the approximation of the formation of a PBH distribution with the same mass for every PBH is favored from the standard mechanism of formation we are going to discuss, whereas in the other cases one should consider a PBH mass spectrum.

After the reheating PBHs may have formed as a result of initial inhomogeneities, regardless of the nature of these peaks in the density distribution.
If the equation of state is hard, that is $p=w\,\varrho$ with $0<w<1$, PBHs forming at time $t_{f}$ from inhomogeneities must have an initial mass of order the particle horizon mass \cite{bjcarr}:
\be
\label{Mhor}
M_{i} \approx M_{H}(t_{f})\approx m_{Pl}^{2}\,t_{f}\approx 10^{5}(t_{f}/s)M_{\odot}\,,
\ee 
that is the overdense regions where PBH formation occurs must have a density contrast $1/3<\delta<1$ (in most models a value closer to one is preferred).   
This is because $M_{i}$ must be bigger than the Jeans mass $\sim w^{3/2}\,M_{H}$ in order to collapse against the pressure but smaller than $M_{H}$ in order not to be in a separate universe. From this general considerations, for example, one has 
\be
\label{tform}
t_{f}=10^{-25}\,s\,, \qquad T\approx 10^{10}\, \textrm{GeV}\,, \qquad M_{i} \approx 10^{18}\,m_{Pl}\,.
\ee

The probability of the PBH formation on any mass scale depends on the amplitude of the density fluctuations on that scale when it enters the particle horizon. Usually a Gaussian window function is imposed in the spectrum mass to calculate the PBH formation probability. The mass spectrum of PBHs has been calculated in different models, and a treatment of this subject is beyond the scope of this work. Let us only mention that the scale invariant  PBH mass spectrum for $M_{i}>M_{cr}$ is $dn/dM \propto M^{-5/2}$ and that different corrections to this power law have been proposed depending on the models,  
whereas in the context of double inflationary models another kind of mass spectrum has been considered, due to the near critical gravitational collapse (see e.g. \cite{spectrum2} and references therein). In the next section we will assume that all PBHs formed at the same time with the same initial mass.

PBHs forming via phase transitions or bubble collisions or the collapse of a cosmic loops might have a smaller mass then the horizon mass, but their production would occur only during a limited period of time and hence over a limited mass range. Considering the quark-hadron transition, for example, when $T\approx \Lambda_{QCD}\,\approx 300$ GeV, the decrease of the pressure forces contrasting the gravitational collapse could have formed a lot of PBHs of mass of the order $1 M_{\odot}$ \cite{Machos} (this idea was considered to address the microlensing observations of MACHOs in that mass range more than one decade ago).

Modifications to the above mentioned formation process and consequent constraints may come from variations of the gravitational constant, as in scalar-tensor theories \cite{Bousso6} or in braneworld models \cite{Bousso7}.

The total energy density in the universe at the time of PBH formation was partitioned between the energy density of radiation $\varrho_{R}$ and PBHs
 $\varrho_{PBH}$: $\varrho(t_{i})=\varrho_{R}(t_{i})+\varrho_{PBH}(t_{i})$, with
\be
\label{part}
\varrho_{R}(t_{i})=(1-\beta)\,\varrho(t_{i})=\frac{\pi^{2}}{30}\,g_{*}\,T^{4}(t_{i})\,,\quad \varrho_{PBH}(t_{i})=\beta\,\varrho(t_{i})=M_{i}\,n_{PBH}\,,
\ee
$\beta$ being a parametrization of the partition, namely the fraction of the universe's mass going in PBHs with the required initial mass.

The first constraint on the PBH number density was based on $\gamma$-rays observations on energies around 100 MeV \cite{HP}. Since then several evidences were collected to improve the limits and to explore a different mass range for the initial PBH masses.
For PBHs with $M_{i}=M_{cr}$, as the observed $\gamma$-ray background density around 100 MeV is $\Omega_{\gamma}\sim 10^{-9}$ and the fraction of the energy emitted from the PBH in form of photons is $\varepsilon_{\gamma} \sim 0.1$, one obtains the bound \cite{boubou}
\be
\beta(M_{cr})<10^{-28}\,,\qquad(M=M_{cr})\,.
\ee

From the entropy production one can derive constraints on $\beta(M)$ for PBHs with initial mass smaller than $M_{cr}$, obtaining \cite{bound11}
\be
\beta(M)<10^{-8}(M/10^{11}\,g)^{-1}\,,\qquad(M<10^{11}\,g)\,.
\ee
From observation on distortions of the CMBR spectrum one gets \cite{b13}
\be
\label{bound13}
\beta(M)<10^{-18}(M/10^{11}\,g)^{-1}\,,\qquad(10^{11}\,g<M<10^{13}\,g)\,.
\ee
From possible variations in the abundances of the primordial elements produced during BBN due to particles emitted by the PBHs one obtains the following bounds:
\be
\beta(M)<10^{-15}(M/10^{9}\,g)^{-1}\,,\qquad(10^{9}\,g<M<10^{13}\,g)
\ee
from the increase of the photon-to-baryon ratio by PBH photons emitted after BBN \cite{b10};
\be
\beta(M)<10^{-21}(M/10^{10}\,g)^{1/2}\,,\qquad(M>10^{10}\,g)
\ee
from the photodissociation of deuterium by PBH photons emitted after BBN \cite{b11};
\be
\beta(M)<10^{-16}(M/10^{9}\,g)^{-1/2}\,,\qquad(10^{9}\,g<M<10^{10}\,g)
\ee
from the modification of the neutron-to-proton ratio by PBH nucleons emitted before BBN \cite{b12}.

These constraints can be modified from the consideration that BHs emit elementary particles. Therefore when the PBH temperature reached values below the threshold $T\approx \Lambda_{QCD}\,\approx 300$ GeV the emitted particles are no longer nucleons, but jets of hadrons. The subsequent hadronization of the jets alters the impact of the PBH evaporation on BBN abundances \cite{hadro}. As decades have passed since the first bounds have been proposed, new measurements and refined theoretical models constantly improve the limits on PBH number density (see e.g. \cite{Kh15}), as is the case for antiprotons and antideuterons (see e.g. refs. \cite{pano}).

There is also a limit for stable Planck mass remnants of PBHs \cite{Bousso6}:
\be
\beta(M)<10^{-27}(M/10^{-5}\,g)^{3/2}\,,\qquad(M<10^{11}\,m_{Pl}\sim 10^{6}\,g)\,.
\ee
All these constraints can be modified if the equation of state in the early universe is ever softer.

Another chance of detection of PBHs stands in neutrinos from the last explosive phase of the PBH emission \cite{neupbh}. Future neutrino telescopes could help to improve the current bounds.

The behavior of $ \varrho_{PBH} \sim a^{-3}$ and of $ \varrho_{R} \sim a^{-4}$, together with the above mentioned observational constraints, lead to conclude that PBHs never dominated the energy density of the universe.

\section{Electrogenesis models}

In discussing these models, different situations can be realized: one can deal with a constant photon mass, which remains the same during all the universe history, or with a temporary photon mass acquired via spontaneous breaking of $U(1)$ electromagnetic gauge symmetry and its later restoration. 
The latter symmetry cannot remain spontaneously broken, otherwise it would lead to the existence of a new light scalar, which is excluded by experiments \cite{Voloshin}.

Another distinction can be done between the case of electric charge conservation and electrogenesis mediated by asymmetric decays of heavy particles  (analogous to the above mentioned baryogenesis scenario), and the case of electrogenesis via PBH disappearance without considering heavy particles decays. 

We will also consider both the cases of charged PBHs and uncharged PBHs, the latter situation proceding in the same way, regardless of the possibility of the critical electric field for the BH Coulomb field and consequent pair production, as will be discussed in the following.

Let us start presenting the case of a constant photon mass, which is then subject to bound \eqref{m3}, and consequently is the less favorable scenario.
In order to make fruitful the mechanism of vanishing of Coulomb field after time \eqref{tauC}, from \eqref{m3} we obtain a lower limit for the BH lifetime:
\be
\label{tauex}
m_{\gamma} < 10^{-27} \textrm{eV}\qquad \rightarrow \qquad \tau_{BH}\geq \tau_{C} \geq 10^{4}\, {\rm yrs}\,.
\ee

From \eqref{BHLT}, this bound translates in the requirement of a PBH initial mass $M_{i}$:
\be
\label{Min}
\tau_{BH}\geq \tau_{C} \geq 10^{4} \textrm{yrs}\,\quad \rightarrow \quad M_{i}\geq 10^{18}\,m_{Pl} \,\approx 10^{13}\, g\,.
\ee

With a smaller $\tau_{BH}$, the process, although still present (the vanishing of the BH Coulomb field is a continous process), would be less efficient. 

Let us assume that all PBHs have mass $M_{i}\approx 10^{18}\,m_{Pl}$. The discussion could be extended to PBHs with bigger masses, although the time of effective electrogenesis will depend on their initial mass. The chosen value \eqref{Min} for the BH mass shows that it must be a PBH, because no reasonable mechanism is known to form a non primordial BH with mass of less than some solar masses.

Consider first the situation of a heavy neutral $X^{0}$ particle, emitted by the evaporation of a Schwarzschild PBH, which decays near the horizon into two charged particles (in analogy with the above mentioned mechanism for baryogenesis via heavy particles decays near a BH \cite{DolgovHid}): 
\be
X^{0} \rightarrow A^{+}\,B^{-} \qquad \textrm{and} \qquad \overline{X}^{0} \rightarrow A^{-}\,B^{+}\,.
\ee
If C and CP are broken in X-decays, then the branching ratios of decays in particles and antiparticles are different: 
\be
BR(X^{0} \rightarrow A^{+}\,B^{-}) \neq BR(\overline{X}^{0} \rightarrow A^{-}\,B^{+})\,.
\ee 
As long as gravitational attraction is stronger than electric repulsion the electric charge may be accumulated inside the BH, because the back-capture of particles of higher mass, say $m_{A} > m_{B} $, is more probable than the one of particles with smaller mass. 
For example the neutral heavy particle could decay in a heavy $t$ quark of mass $m_{t}$ and a light $\overline{u}$ antiquark of mass $m_{u}$, once during evaporation the condition $T_{BH}> m_{t}$ is reached. 

This process may work even with massless photons, but the accumulation of charge in the PBH must stop when the Coulomb force becomes equal to the gravitational one. In this case the maximum accumulated charge (in multiples of unitary charge $e$) is:
\be
\label{36}
\epsilon = \frac{N_{charge}}{N_{PBH}}=\frac{m^{2}_{p}}{\alpha\,m^{2}_{Pl}}=10^{-36}\,,
\ee  
where $N_{charge}$ is the number of excessive unitary charge, $N_{PBH}$ is the number of protons forming the PBH (it is simply a way to parametrize the PBH mass), $\alpha$ is the electromagnetic coupling constant and $m_{p}$ is the mass of the proton.

This is the classical limit found in the standard mechanism proposed to charge astrophysical compact objects  \cite{Schwarzman}. The reason why we need a massive photon stands in the further requirement of the PBH Coulomb field disappearance. The latter process is necessary in generating electrogenesis to overcome the obstacles of electric charge conservation and the existence of the electric hair of the PBH.

Considering massive photons, instead, after a time interval $\Delta\,t \approx 1/m_{\gamma}$ the electric field vanishes: this process effectively represents a disappearance of the BH electric charge. After $\Delta\,t$ the PBH can start from a new initial electrical neutral state to accumulate charge. With the indicated value for $M_{i}$, the PBH Coulomb field can vanish completely, without suppression in the efficiency of the mechanism, as is shown in \eqref{tauex}. Thus, also imposing the very stringent bound \eqref{m3} on the photon mass, 
one obtains the generation of an uncompensated electric charge $\epsilon$ by every PBH. The choice of the initial value $M_{PBH}=10^{18}\,m_{Pl}$ means that every PBH can possess, immediately after the Coulomb field disappearance, a charge $Q=10$.

Exactly the same result would be obtained considering the asymmetric decay of a charged heavy particles, or of a neutral heavy particle emitted by a charged PBH or by both variations to the presented situation, because it relies on the limit \eqref{36}, which is unaffected by those differences.  

As the evaporation is an explosive process, during the last stage of the BH life-time very heavy particles can be produced by virtue of \eqref{BHT}, regardless of the temperature of the universe at that time. By the same reason, the higher is the temperature of the BH, the smaller is the time interval it keeps this value of the temperature. During most of the BH life-time no heavy particles will be produced by evaporation. If the initial BH density is small enough, the particles escaped for the BH would not affect significantly the CMBR spectrum.

From observations we know that during BBN $|\Delta| \leq 10^{-32} e$. Assuming that all PBHs have the same $M_{i}$ and the same electric charge $Q$ (expressed in multiples of the unitary charge $e$), every PBH should possess no more then a charge $Q=10$. As $\Delta$ represents the ratio of the electric charge density over the  baryonic charge density, we can get a value for the ratio of PBH density over baryon density:
\be
\label{nBH}
\frac{n_{Q}}{n_{B}} \leq 10^{-32} \qquad \rightarrow \qquad \frac{n_{PBH}}{n_{B}} \leq 10^{-33}\,.
\ee  

Applying the same reasoning with the less restrictive bound $|\Delta|<10^{-29} e$, one can increase the previous limit only by three orders of magnitude. From the previous section we can compare this result with the observational and theoretical constraints on PBH number density, to see how reasonable can be these assumptions on the initial PBH distribution.
 
Let us estimate $n_{PBH}$ at the time when in the universe $T \approx 40 MeV$, that is before BBN and just before the protons decouple from the photons, which allows one to use the equilibrium distribution for the density of the protons.

The condition of decoupling is that the decay rate $\Gamma$ of the protons (through processes mediated by electromagnetic interactions with photons) becomes comparable to the Hubble radius $H$: $\Gamma \approx H$. Considering a flat universe, from the Friedmann equations one has 
\be
H^{2}=\frac{8\,\pi}{3\,m^{2}_{Pl}}\,\varrho\,.
\ee
In the case of a radiation dominated universe, one can write
\be
\varrho \approx \varrho_{R}=\frac{\pi^{2}}{30}\,g_{*}\,T^{4}\,.
\ee
From the following identities
\be
\label{np}
\Gamma \approx \sigma_{T}\,n_{B}\quad,\quad\sigma_{T}=(8\,\pi/3)(e^{4}/m_{p}^{2}) \approx 10^{-31} {\rm cm}^{2}\quad,\quad n_{B}=g_{*}\,\left(\frac{m_{p}\,T}{2\,\pi}\right)^{3/2}\,e^{-m_{p}/T}  \,, 
\ee
where $\sigma_{T}$ is the Thompson cross section for protons, and $n_{B}$ in this case is the protons equilibrium distribution with negligible chemical potential, one gets 
\be
\frac{{\rm GeV}}{T}+\frac{1}{2}\,\ln\,(T/{\rm GeV}) \approx \ln 10^{18} \quad \rightarrow \quad T \approx  40\,\textrm{MeV}\,.
\ee
Putting this value in the distribution \eqref{np} gives the estimate
\be
\label{estnb}
n_{B} \approx 10^{-13} {\rm GeV}^{3}\,\approx 10^{14} {\rm eV}^{3}\,\approx 10^{28} {\rm cm}^{-3}\,.
\ee

The presence of the logarithms in the previous estimate may change considerably the value $T \approx 40 MeV$ for small variations, e.g. in $g_{*}$. One can have another confirmation from the following considerations. Ignoring changes in the entropy density due to particle decouplings, one has
\be
n_{B}\approx 10^{-9}\,s\quad,\quad s=\frac{2\,\pi^{2}}{45}\,g_{*}\,T^{3}\,,
\ee
where $s$ is the entropy density.
Inserting $T=40$ MeV one obtains once again the value \eqref{estnb} $n_{B} \approx 10^{-13} {\rm GeV}^{3}$. 

Taking $M_{i}=10^{18}\,m_{Pl}$, $n_{PBH}=10^{-33}\,n_{B}$ and $g_{*} \approx 20$, one gets $n_{PBH} (T=40 MeV) \approx 10^{-33}\,10^{28} cm^{-3}=10^{-5}  cm^{-3}$. 

We can now compare this result with the constraints of the previous section. As both at the time of the PBHs formation $t_{f}\approx 10^{-25}$ s and at the time $t_{40}\approx 10^{-3}$ s discussed here the universe was in the radiation dominated era, we can write
\be
\frac{n_{i}}{n_{40}}\,=\,\left(\frac{t_{f}}{t_{40}}\right)^{-3/2}\qquad\rightarrow\qquad n_{i}=10^{28}\,{\rm cm}^{-3}\,.
\ee

Bound \eqref{bound13} $\beta \leq 10^{-20}$, together with the conditions \eqref{tform} and the identity \eqref{part}, give an initial number density for PBHs $n_{t_{f}}\approx 10^{25}\,{\rm cm}^{-3}$. 

This comparison indicate that, among all the initial PBHs formed at the same time $t_{f}$ only a small fraction of them can get an electric charge $Q$.
Bounds \eqref{bound13}, in fact, impose that the previous assumption \eqref{nBH} must be changed in the following way:
\be
\label{nBH2}
\frac{n_{Q}}{n_{B}} \leq 10^{-32} \qquad \rightarrow \qquad \frac{n_{PBHC}}{n_{B}} \leq \varepsilon \cdot 10^{-33}\,,\qquad \varepsilon =10^{-3}\,,
\ee  
where $\varepsilon=10^{-3}$ is the fraction of charged PBHs with respect to their total initial distribution.

This suppression regards the number density of PBH with the required initial mass assuming that the only PBH that formed have that mass. If one adopts a mass spectrum, then the constraints on $\Delta$ in that case regard all PBHs, so the number density of the PBHs with a certain initial mass should be lower than the value we found.

To get a deeper understanding of the correctness of the obtained value for $Q$, we need to discuss some more properties of the charged BHs \cite{FN}, which will be useful also for the other realizations of electrogenesis via PBHs. As these properties may concern quantum gravitational effects and the final role of the BH, it is notewhorty to mention the following statement about the interior of a BH.
  
It was shown \cite{CH} that, in general, all global charges are extinguished before the infalling matter crosses the singularity by the wormhole-induced global-charge violation mechanism. If the photon is massive, regardless of the nature of their mass, the electric charge becomes a global charge. Therefore, this result may have some importance for deep analysis of our models of electrogenesis.

An effect which probably diminishes the BH charge $Q$ is due to the possibility of the presence of a critical electric field $E_{cr}$. Once this value is reached, the pair production becomes possible. To estimate $E_{cr}$ one can equate the electric field $E=Q/r^{2}$ at the horizon $r=R_{g}=2M_{BH}/m_{PL}^{2}$ (which is where mainly the process occurs) to the value  $E \approx m_{e}^{2}$, where $m_{e}$ is the electron mass. This means that when the BH charge $Q$ reaches the value
\be
\label{Ecr} 
Q=\frac{m_{e}^{2}\,M_{BH}^{2}}{m_{Pl}^{4}}\,
\ee
 pair production will take place and $Q$ will be strongly compensated by the particles produced in this way. If $Q$ is such as to generate an electric field much lower than the critical one, than instabilities of the BH and true singularities will be avoided \cite{FN}.

It was also shown that isolated BHs with mass $M_{BH}\leq  10^{15} g$ lose their electric charge almost completely and very rapidly \cite{Ruf}.  

Whereas an uncharged BH can evaporate without serious problems until its complete disappearance, because of the  absence of hairs connected with the baryonic charge, a charged BH can not evaporate away completely. This statement can be faced from different points of view. 
The fact that the BH temperature goes to zero in the Reissner--Nordstrom extreme limit $Q^{2}=(M_{BH}/m_{Pl})^{2}$ indicates that the evaporation slows down and does not proceed past the point $Q^{2}=(M_{BH}/m_{Pl})^{2}$ (see e.g. \cite{UVIR}). This extreme limit can be viewed as the ground state of the charged BH. Apart from this quantum mechanical consideration, also in classical field theory a charged object cannot go below this limit \cite{FN}.

Therefore, from these different perspectives it seems very reasonable to suppose that after evaporation a PBH will stop the emission and remain in the above mentioned extreme limit. With our value of $M_{i}$ this limit would be $Q=10^{18}$. In this case the contribution of such PBHs to the energy density of the universe is strongly suppressed: $n_{PBH}\leq 10^{-51} \cdot n_{B}$.

To resume, it could seem that, for the chosen value of $M_{i}$, an initially charged BH cannot acquire a charge greater then $Q \approx 10$, whereas if it was formed with an its own original electric charge, this charge cannot go below $Q \approx 10^{18}$.

This apparent paradox is easily overcome, once one realizes that in the relation coming from the extreme limit, occurring at a certain point of the evaporation, one cannot insert $M_{i}$, but a certain $M_{BH}(t)$. Knowing that the BH mass decreases in time during evaporation, one can be convinced on the validity of the expression $Q_{max}=10$. 

The pair production due to the reaching of the critical electric field value in our case should diminish the effective value of the PBH charge. Therefore, it seems reasonable to consider that any PBH could possess a final electric charge $Q \approx 1$, after the decreasing due to the pair production. In this case the stable charged remnant of the PBH would be an extremely compact object with unitary electric charge and with a Planck mass. Indeed, observational constraints \eqref{bound13} (with the further suppression due to $\varepsilon$) exclude these remnants from being good dark matter candidates, nevertheless their existence could give the universe a net electric charge.

Considerations on stability, in fact, are based on the persistence of the PBH Coulomb field. The disappearance of the latter allows the remnant to complete its evaporation. Therefore, also if quantum gravity effects are present, but they do not spoil the classical description of the Coulomb field seen by an observer far outside the PBH horizon, a net electric charge of the universe might be created by this mechanism.     

To conclude the discussion, if after asymmetric heavy particle decays a PBH can acquire a certain electric charge $Q$, and if its life-time is bigger than $10^{4}$ years, than this $Q$ can vanish leaving the universe with a net electric asymmetry, represented by the non-compensated electric charges emitted by the PBH evaporation.

The same conclusion can be drawn for an initially isolated charged PBH of initial mass \eqref{Min}, ignoring the heavy particle decays. Through pair production its original electric charge will decrease and after a time $\tau_{C} \approx 10^{4}$ years the electric asymmetry will be created.

The last situation we are going to discuss is the possibility of a spontaneous breaking of the electromagnetic gauge invariance and its later restoration. This feature has been previously considered as a source of local electric currents, which may create magnetic fields acting as seeds of cosmic galactic magnetic fields \cite{DS}. In the mentioned work the electrogenesis is not considered because, when the symmetry is broken, the same but opposite electric charge asymmetry is developed in the vacuum of the theory. When the symmetry is restored, the charge of the vacuum will reappear in the form of heavy charged Higgs bosons and the net electric charge of the universe must be zero.

Nevertheless, we want to consider the opportunity of restoration of the spontaneous symmetry breaking just after PBHs evaporation. If the symmetry is restored after the Coulomb field of PBH has vanished, the electric charge carried by the heavy Higgs bosons will be no more compensated by the charged particles which crossed the PBH horizon. In this way the electrogenesis can be created by this charged heavy Higgs bosons at the moment of the  gauge symmetry restoration. The decay products of the heavy particles could not distort much the CMBR spectrum with appropriate values for the parameters of the transition. 

Moreover, in the case of spontaneous breaking of the electromagnetic $U(1)$ gauge invariance by the coupling $\mathcal{L}_{int}= \xi R A^{\mu}A_{\mu}$, bounds \eqref{m1}, \eqref{m2} and \eqref{m3} do not apply to the photon mass, because they are obtained only under the assumption of a constant mass of the photons. This in turn allow one to take considerably smaller $\tau_{C}$ and hence smaller $\tau_{BH}$ as discussed in ref. \cite{DolgovMe}.
The consequent generation of electric charge in the universe may be bigger than the previously discussed one.

Another interesting feature of this kind of electrogenesis stands in the possibility of having at the same time a non null electric charge density $\sigma$ and a standard electrodynamics with massless photons after the restoration of the symmetry. Maxwell equations in this case, as already mentioned, give rise to an electric field rising with distance, which destroys the homogeneity and the isotropy of the Friedmann--Robertson--Walker metric. The formation of domains in the universe might happen in this case, each domain having as its size the horizon size at the time of the symmetry restoration. If the differents domains had their electric fields pointing along different directions, as result of a stochastically distributed orientation of the electric fields in all the domains, in the regions near the domain walls interesting features might be present. 

The energy release along the domain walls might create strings of overdensity in the matter distribution, as the filaments we observe in the large scale structures. The size of the domains in the present sky would depend on their original size, that is on the time of the symmetry restoration. Therefore any prediction about the observable consequences of these phase transitions woul be strongly model-dependent.

In conclusion, we have presented different possible realizations of electrogenesis at an epoch between BBN and last scattering, which rely on the horizon properties of PBHs and on the vanishing of the PBH Coulomb potential in the case of massive photons.

\section{Conclusions}

Some estimates found that a net electric charge in the universe is efficiently driven away in a time small compared to the Hubble time for temperatures $100$ GeV$ \geq$ T $\geq 1$ eV \cite{Siegel}. Nevertheless, in our mechanism the evaporation process of PBHs must last for $10^{4}$ years in order to satisfy the stronger bound on the photon mass \eqref{m3}. Electrogenesis takes place only after the Coulomb field of the PBHs has disappeared, that is only when the temperature in the universe is well below the value $T \approx 1$ eV. The initial conditions required to PBHs allows the generated electric asymmetry to avoid the predicted washing out.

We discussed a mechanism which gave a net electric charge to the universe in the past. The same considerations can be applied to PBHs with higher initial mass, which have not yet evaporated. Provided the appropriate observational constraints are imposed on their density from the above mentioned different limits for the different mass ranges, one can estimate in the same way as we did the net electric charge which might be created in the near future. Moreover, PBHs with higher masses would not be subject to the above mentioned washing out of the electric charge.

Among different realisations of the mechanism, the more interesting seems to be the last one, that is the model in which the mass of the photon comes from a spontaneous symmetry breaking and the electric charge asymmetry comes from a later symmetry restoration. A detailed analysis of the dependence of the results from the parameters of the symmetry breaking model and their possible consequences in the present universe could be worth.

Wheter or not quantum gravity effects stop the evaporation of the PBHs, once they acquired an electric charge, our mechanism works the same, provided the classical description of the PBH Coulomb field seen by an observer far outside the horizon is not affected by them. 
Another possibility is that these effects could modify the characteristic time of disappearance of the Coulomb field, because the latter has been calculated taking into account the quantum mechanical description of the photons near the BH horizon, where quantum gravitational effects might be very strong.
The situation would be completely unclear in the case of the coupling of the electromagnetic vector potential to the scale curvature, because we have no idea of how the quantum gravitational effects might change the dynamics of the electromagnetic field.

Although experimental limits on $\Delta$ are made at BBN epoch and at last scattering epoch (and at present time), assuming the conservation of electric charge, the cosmological electric asymmetry may be a small number. Nevertheless, the large scale properties of a mild charged universe can present interesting differences from the ones of a neutral universe.

All different models of electrogenesis via PBHs that we showed lead to a small but non vanishing value of the net electric charge of the universe. Further investigations devoted to a deeper understanding of the impact of various scenarios and relative consequences on large scales may be very interesting.

\acknowledgments

I am very grateful to A.Dolgov for precious suggestions and comments. I also wish to thank F.Urban for helpful discussions.

\end{document}